\documentstyle[twocolumn,aps,prl,epsf,epsfig]{revtex}
\begin{document}
\draft
\title{Entangling Two Bose-Einstein Condensates by Stimulated Bragg Scattering}
\author{B. Deb and G.S. Agarwal}
\address{Physical Research Laboratory, Navrangpura, Ahmedabad-380009, India}
\date{\today}
\maketitle
\begin{abstract}
We propose an experiment for entangling  two spatially separated Bose-Einstein
condensates by  Bragg scattering of light.  When Bragg scattering in two
condensates is stimulated by a common  probe, the resulting quasiparticles in
the two condensates get entangled due to quantum communication between the
condensates via probe beam. The entanglement is shown to be significant and
occurs in both number and quadrature phase variables. We present two methods of
detecting the generated entanglement.
\end{abstract}
Inseparability of quantum states of two or more subsystems is the most
significant feature of quantum mechanics. Apparently puzzling, yet most
profound, first formulated as a paradox \cite{epr}, this inseparability known
as quantum entanglement lies at the very heart of nonclassical physics.
Further, as a basic resource for quantum information processing, it has become
a focal theme of research in modern physics and many issues in the foundations
of quantum mechanics. Generation and manipulation of entanglement is,
therefore, of prime  interest.  Bose-Einstein condensates (BEC) \cite{bec} of
weakly interacting atomic gases seem to be suitable macroscopic objects for
producing many-particle  entanglement  \cite{sorensen}.  A BEC has intrinsic
entanglement character  due to reduced quantum fluctuations in momentum 
space.  For instance, in the condensate ground state,  a pair of mutually
opposite momentum modes is maximally entangled in atomic number variables
\cite{deb}. 

Stimuated resonant Bragg scattering of light by a condensate generates
quasiparticles \cite{quasi},  predominantly in two momentum side-modes
$\mathbf{q}$ and $-\mathbf{q}$, where $\mathbf{q}$ is the momentum transfered
from light fields to the atoms. Momentum side-modes are the excited states of a
BEC, atoms in such a state collectively behave as quasiparticles. Bragg
spectroscopy \cite{bragg} with coherent or classical light produces   coherent
states of the quasiparticles in a BEC. When these quasiparticles are projected
into particle domain, that is, into the Bogoliubov-transformed momentum modes
\cite{bogoliubov}, they form two-mode squeezed as well as entangled state
\cite{deb}.  Bragg spectroscopy  with nonclassical light can  generate 
tripartite entanlement \cite{deb} in a condensate.  In addition to atom number
and phase variables, spin degree-of-freedom of a spinor BEC \cite{spinor} can 
be useful in describing entanglement in spin variables. Thus, BECs offer a
fertile ground  for studying different aspects of entanglement.  Apart from
BECs, multi-atom entanglement in other  macroscopic systems has been realized 
\cite{spinexp} on the basis of collective spin squeezing \cite{spin1,spin2}. 
Further the  entanglement in collective spin variables of two ensembles  of
gaseous Cs atoms  has been experimentally demonstrated \cite{sasha} .
Continuous variables like the quadratures of a field  mode (which are 
analogous to  position and momentum) have also been employed \cite{kimble} in
entanglement studies. 

We here propose a scheme for producing quantum entanglement between two
spatially separated BECs of a weakly interacting  atomic gas.  The entanglement
we consider is in  quasiparticles  of BECs.    The  proposed experiment is
schematically shown in Fig.1. The condensates A and B  are illuminated  by pump
lasers L1 and L2, respectively. A single stimulating probe laser L3 passes
through both the condensates. All these three lasers are detuned far off the
resonance of an electronic excited state of the atoms. The frequencies and the
directions of propagation of these lasers  are so chosen  such that Bragg
resonance (phase matching) conditions of scattering  in both  the condensates
are fulfilled. The Hamiltonian of the system is $H=H_{A}+H_{B}
+H_{F}+H_{AF}+H_{BF}$. Retaining the dominant momentum side-modes $\mathbf{q}$
and -$\mathbf{q}$ only under Bragg resonance condition, in the Bogoliubov
approximation \cite{bogoliubov}, $H_A = \hbar \omega_{q}^B
\left(\hat{\alpha}_{{\mathbf{q}}}^{\dagger} \hat{\alpha}_{{\mathbf{q}}}
+\hat{\alpha}_{-{\mathbf{q}}}^{\dagger} \hat{\alpha}_{-{\mathbf{q}}}\right)$,
where $\hat{\alpha}_{\mathbf{q}}$ represents quasi-particle with mometum
$\mathbf{q}$, and  $\omega_{q}^B=\left[(\omega_q + \frac{\mu}{\hbar})^2 -
(\frac{\mu}{\hbar})^2\right]^{1/2}$ is the frequency  of Bogoliubov's 
quasi-particle \cite{bogoliubov}.  Here $\omega_q = \frac{\hbar^2q^2}{2m}$,
and  $\mu = \frac{\hbar^2\xi^{-2}}{2m}$ is the chemical potential with $\xi =
(8\pi n_{0}a_{s})^{-1/2}$ being the healing length. Similarly, $H_B = \hbar
\omega_{q}^B \left(\hat{\beta}_{{\mathbf{q}}}^{\dagger}
\hat{\beta}_{{\mathbf{q}}} +\hat{\beta}_{-{\mathbf{q}}}^{\dagger}
\hat{\beta}_{-{\mathbf{q}}}\right)$, 
\begin{figure}
\psfig{file=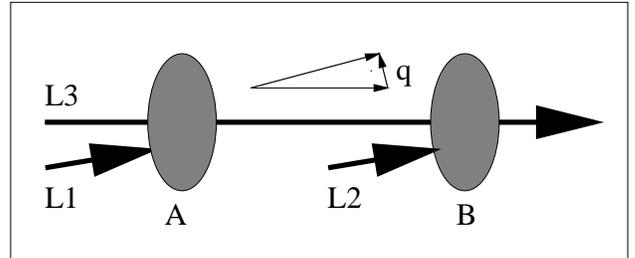,width=3.25in}
\caption{ The scheme for creation of entanglement. A and B are two
condensates, L1 and L2 are pump lasers, L3 is a common entangling  probe laser.
Both the pumps have same wave vector ${\mathbf{k}}_1$, probe's wave vector is
${\mathbf{k}}_2$. The probe is red-detuned from the pumps. The lasers are in
Bragg resonance with a particular  momentum mode $\mathbf{q}$ of both the
condensates.}
\label{fig1}
\end{figure}
wih $\hat{\beta}_{\mathbf{q}}$ being the quasi-particle operator of the
condensate B.  The pumps are treated classically. Let  $\hat{c}$ represents the
common probe field mode, then the field Hamiltonian $H_F = -\hbar \delta
\hat{c}^{\dagger}\hat{c}$, where  $\delta = \omega_1 - \omega_2$ is the
pump-probe detuning.   The quasiparticle operators $\hat{\alpha}(\hat{\beta})$
are related to the particle operators  $\hat{a}(\hat{b})$ by Bogoliubov's
transformation : $\hat{a}_{{\mathbf{q}}}=
u_{q}\hat{\alpha}_{\mathbf{q}}-v_{q}\hat{\alpha}_{-\mathbf{q}}^{\dagger}$,
where $v_{q} = (u_{q}^{2}-1)^{1/2} = [\frac{1}{2}(\frac{\omega_q +
\mu/\hbar}{\omega_{q}^B}-1)]^{1/2}$.   The atom-field ineteraction Hamiltonian
for condensate A is
\begin{eqnarray}
H_{AF} = \hbar\eta_{A}
\hat{c}^{\dagger}
(\hat{\alpha}_{\mathbf{q}}^{\dagger} + \hat{\alpha}_{-{\mathbf{q}}})
+{\mathrm H.c.}
\end{eqnarray}
where $\eta_A = \sqrt{N_A}\Omega_A f_q$ is the effecive atom-field coupling
constant. Here $N_A$ is the number of atoms in condensate A , $\Omega_A$ is the
two-photon Rabi frequency of an atom in A and $f_q = u_q - v_q$. $H_{BF}$ is
given by the similar expression as $H_{AF}$ with subscript $A$ replaced by $B$
and $\alpha$ replaced by $\beta$. The Heisenberg equations of motion are
\begin{eqnarray}
\dot{\hat{\alpha}}_{\mathbf{q}} = -i\omega_q\hat{\alpha}_{\mathbf{q}} 
-i\eta_A \hat{c}^{\dagger}
\end{eqnarray}
\begin{equation}
\dot{\hat{\alpha}}_{-\mathbf{q}}^{\dagger} =
i\omega_q\hat{\alpha}_{-\mathbf{q}}^{\dagger} + i\eta_{A}\hat{c}^{\dagger}
\end{equation}
\begin{eqnarray}
\dot{\hat{c}}^{\dagger} = -i\delta \hat{c}^{\dagger} + i[\eta_A
(\hat{\alpha}_{\mathbf{q}} +
\hat{\alpha}_{-\mathbf{q}}^{\dagger}) + \eta_B (\hat{\beta}_{\mathbf{q}} +
\hat{\beta}_{-\mathbf{q}}^{\dagger})]
\end{eqnarray}
The Heisenberg equations of $\hat{\beta}_{\mathbf{q}}$ and
$\hat{\beta}_{-\mathbf{q}}^{\dagger}$ are similar to  those of $\hat{\alpha}$,
but $\hat{\alpha}$ and $\eta_A$ should be replaced by $\hat{\beta}$ and
$\eta_B$, respectively.

We next discuss how to quantify entanglement between two  BECs. If the
entanglement occurs in number operators of the quasiparticle  modes 1 and 2, 
then it can be quantified by the parameter \cite{spin1,deb}
\begin{equation}
\xi_n(1,2) = \langle[\Delta(\hat{n}_{1}-\hat{n}_{2})]^2\rangle/
(\langle\hat{n}_{1}\rangle + \langle\hat{n}_{2}\rangle).
\end{equation}
If $\xi_n < 1$, then the two modes are entangled.  If the entanglement is
described by two noncommuting Gaussian operators $\hat{X}$ and $\hat{P}$ which
are analogous to position and momentum variables, then  the entanglement 
parameter is defined by \cite{xip}
\begin{equation}
\xi_p(1,2) = \frac{1}{2}[\langle[\Delta(X_{1}+X_{2})]^2\rangle
+\langle[\Delta(P_{1}-P_{2})]^2\rangle]
\end{equation}
The two modes are entangled in quadrature phase, when  $\xi_p<1$.

For numerical illustration, we consider two homogeneous identical  Na
condensates.  We here enlist the important results: 1) If the modes
${\mathbf{q}}_1$ of A and ${\mathbf{q}}_2$ of B are in Bragg-resonance with the
the respective Bragg pulses, and if the effective coupling of B ($\eta_B$)  is
stronger than that of A, then entanglement arises between ${\mathbf{q}}_1$ of A
and $-{\mathbf{q}}_2$ of B only,  other pairs of modes are immune to any
entanglement. In fig.2,  we display entanglement parameters between these two
chosen modes  as a function of time. We set
${\mathbf{q}}_1={\mathbf{q}}_2=\mathbf{q}$. The effective coupling can be
\begin{figure}
\psfig{file=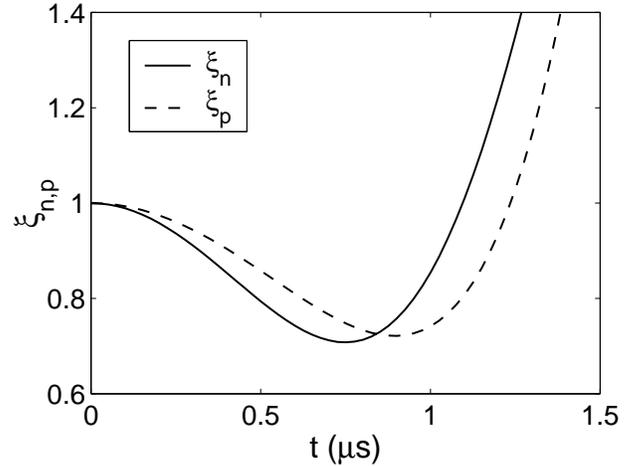,width=3.25in}
\caption{ Entanglement parameters $\xi_n$ and $\xi_p$ between 
${\mathbf{q}}$-mode of A and -$\mathbf{q}$-mode of B as a function of time in
$\mu$s.  For both the condensates, $\omega_q^B = 0.21$ MHz, $q = 2 \xi^{-1}$
and  $\delta=0.17$ MHz. The coupling constants $\eta_A =1.62$ MHz, $\eta_B = 
1.25 \eta_1$. Both the condensates are initially in the ground states  (vacuum
of quasi-particle operators), and the common probe field is in coherent state
with average number of photons equal to 10.}
\end{figure}
made different either by using pump lights of different intensities or taking
different atom numbers for the two otherwise identical condensates. 2) For
equal couplings, there is no entanglement between any pair of modes.  Fig.3 
shows the variation of entanglement parameters as a function of the ratio of
the two coupling constants at a fixed time.  3) We find entanglement both in
quasiparticle (phonon) modes ($\hat{\alpha},\hat{\beta}$), and in the
Bogoliubov transformed modes of quasiparticles  which we call particle or 
atomic modes ($\hat{a},\hat{b}$). However, in  atomic modes, entanglement is
weaker than that in quasiparticle modes.  It is worth mentioning that in a
single condensate, as shown in Ref.\cite{deb}, coherent light scattering can
generate entanglement only in atomic modes, and not in phonon modes. In
contrast, one can generate {\it entanglement in phonon modes in two
condensates} by coherent light scattering.    The light scattering events
occuring at  A and B are not independent, since a quantum communication has
been set between the generated quasiparticles in A and B via the common probe. 
Had we treated the common {\it probe classically},  then the Hamiltonian 
(Eq.(1))  would have been  linear in the atomic operators. A Hamiltonian linear
in Bosonic opeartors  can not generate nonclassical correlation. Therefore, the
{\it probe} must be treated {\it quantum mechanically}. The probe carries with
it quantum fluctuations  of one condensate and transfers a part of it to the
other leading to the entanglement  between the two  condensates.

To explain the results further, we here resort to an approximate analysis. Let
us suppose, $\omega_q <\!<\eta_{A(B)}$ and $\delta <\!<\eta_{A(B)}$, then we
can neglect the diagonal terms proportional to $\omega_q$ and $\delta$ in the
Hamiltonian. For equal 
\begin{figure}
\psfig{file=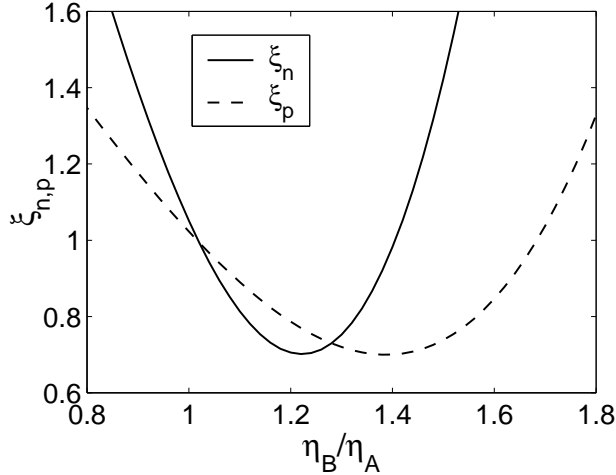,width=3.25in}
\caption{  The entanglement parameters $\xi_n$ (solid) and $\xi_p$ (dashed) as
a function of $\eta_B/\eta_A $ for a fixed  time $t=0.75$ $\mu$s. The other
parameters are the same as in Fig.2.}
\end{figure}
coupling ($\eta_A = \eta_B$), from  Heisenberg equation of motions,
it then follows that  $\hat{\alpha}_{\mathbf{q}}(t) +
\hat{\beta}_{-\mathbf{q}}^{\dagger}(t) = \hat{\alpha}_{\mathbf{q}}(0) +
\hat{\beta}_{-\mathbf{q}}^{\dagger}(0)$, that is, the superposition operator
$\Sigma = \hat{\alpha}_{\mathbf{q}} + \hat{\beta}_{-\mathbf{q}}^{\dagger}$
becomes a constant of motion. Let us write the quadratures $X_A =
\frac{1}{\sqrt{2}}(\hat{\alpha}_{\mathbf{q}}  +
\hat{\alpha}_{\mathbf{q}}^{\dagger})$, $P_{A} = 
-i\frac{1}{\sqrt{2}}(\hat{\alpha}_{\mathbf{q}} - 
\hat{\alpha}_{\mathbf{q}}^{\dagger})$, and similarly for $X_B$ and $P_B$. Then
it can be shown that $\xi_p = \frac{1}{4}[\langle[\Delta(\Sigma +
\Sigma^{\dagger})]^2\rangle -
\langle[\Delta(\Sigma-\Sigma^{\dagger})]^2\rangle]$. For equal coupling and the
initial states being  in vacuum or in coherent states, one obtains $\xi_p =1$,
that is, the two modes are  unentangled. Let us then consider the case  of
different couplings, for short times characterized by $\eta_At <\!<1$, and 
$\eta_Bt <\!<1$, we obtain perturbative solutions of
$\hat{\alpha}_{{\mathbf{q}}}(t)$ and $\hat{\beta}_{-{\mathbf{q}}}(t)$  upto the
second order in time. Using these solutions, we calculate $\xi_p = 1 -
\eta_A\eta_Bt^2(1 - \frac{\eta_A}{\eta_B})$, which is less than unity (the two
modes are entangled in quadrature variables) if $\eta_A\eta_Bt^2(1 -
\frac{\eta_A}{\eta_B}) > 0$ which is only possible if $\eta_A \ne \eta_B$ and
$\eta_A < \eta_B$.  Similarly, we can prove that for -$\mathbf{q}$
(off-resonant) of A and $\mathbf{q}$ (resonant) of B, $\xi_p = 1 +
(\eta_Bt)^2(1-\frac{\eta_A}{\eta_B})$ which is always greter than unity for
$\eta_A < \eta_B$. For the same  resonant $\mathbf{q}$-mode  of A and B, $\xi_p
= 1 + \eta_Bt^2/2 + (\eta_A^2 + \eta_B^2)^2t^4/4$, which is always greater than
unity.  In the same way, we can show that, for the remaining mode-pair 
($-{\mathbf{q}}, -{\mathbf{q}}$),  $\xi_p$ is also  greater than unity.

Next, we prove that, to generate entanglement in number variables ($\xi_n$),
the two coupling parameters should also be different. Substituting $\hat{n}_1 =
\hat{\alpha}_{\mathbf{q}}^{\dagger}\hat{\alpha}_{\mathbf{q}}$ and $\hat{n}_2 =
\hat{\beta}_{-\mathbf{q}}^{\dagger}\hat{\beta}_{-\mathbf{q}}$ in Eq.(5)  and
using the pertubative solutions  we can express $\xi_n= 1-
R/(\langle\hat{n}_{1}\rangle + \langle\hat{n}_{2}\rangle)$ where
\begin{eqnarray}
R = 8\eta_A^2t^4[\eta_B^2 - 2\eta_A^2 + 4 n_p (\eta_B^2 -
\eta_A^2)]
\end{eqnarray}
where $n_p$ is the initial number of photons in the coherent probe beam. Now,
$\xi_n < 1$ implies that $R>0$ which amounts to $(\eta_B/\eta_A)^2 > 1+
1/(1+4n_p)$, that is, $\eta_B > \eta_A$. On the other hand, if $\eta_B\le
\eta_A$, $\xi_n > 1$.  We also carry out  an aletrnative analysis to check 
whether the two resonant modes $\mathbf{q}$ of A and B exhibit any entanglement
in other parameter regimes. By neglecting the off-resonant mode $-\mathbf{q}$
in both the condensates and keeping only the resonant mode, it can be
analytically proved that $\xi_p({\mathbf{q}},{\mathbf{q}}) = 1 + \sinh^2(\eta
t)$ and $\xi_n({\mathbf{q}},{\mathbf{q}}) = 1 +
(1+\frac{n_p}{n_p+1})\frac{(\eta_A^2 - \eta_B^2)^2}{(\eta_A^2 + \eta_B^2)^2}
\sinh^2(\eta t)$, that is, both the  parameters $\xi_p$ and $\xi_n$ are always
greater than unity. Here $\eta = \sqrt{\eta_A^2 + \eta_B^2}$. 

We next show how a set up as shown in Fig.4 can be utilized  to verify the
generated entanglement.  After the process of generation of entanglement, the
duration of which can be typically on the order of 1 to 100 $\mu$s,   is over,
the lasers L1, L2 and L3 are switched off. Two different pairs of verifying
pump-probe Bragg pulses are applied to the condensates, as described in the
caption of Fig.4. The two probes should be derived from a common source. The
modes $\mathbf{q}$ of A and $-\mathbf{q}$ of B are in Bragg resonance with the
respective Bragg pulses. The effective field-condensate couplings for both the
condensates are very small compared to the Bogoliubov frequency $\omega_q^B$.
Let $\hat{c}_{probe,A}$ and $\hat{c}_{probe,B}$ denote the verifying probe
field modes for the condensates A and B, respectively. By neglecting the
off-resonant terms 
$\hat{\alpha}_{2{\mathbf{q}}}^{\dagger}\hat{\alpha}_{{\mathbf{q}}}$ and
$\hat{\alpha}_{-2{\mathbf{q}}}^{\dagger}\hat{\alpha}_{-{\mathbf{q}}}$ in the
Hamiltonian, the time evolution of the output probe modes, in a frame rotating
with pump-probe detuning $\delta$, can be written as
   \begin{eqnarray*}
  \hat{c}_{probe,A}^{(out)} &\simeq&  \hat{c}_{probe,A}^{(in)} +
  \frac{\eta_A}{\delta - \omega_q^B}
  \left(\exp[i(\delta - \omega_q^B)t]-1\right)\hat{\alpha}_{\mathbf{q}}^{\dagger} \nonumber \\
  &+&
  \frac{\eta_1}{\delta + \omega_q^B}
  \left(\exp[i(\delta + \omega_q^B)t]-1\right)\hat{\alpha}_{-\mathbf{q}}
  \end{eqnarray*}
\begin{figure}
\psfig{file=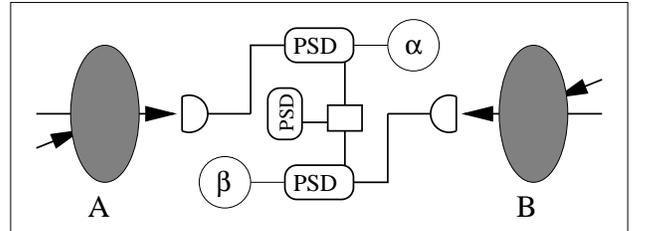,width=3.25in}
\caption{  The scheme   for  verification of the entanglement.  Apart from pump
lasers, two extremely weak verifying probes of the same frequency as that of
entangling probe - one each for each condensate - are switched on . The  probes
acting on A and B have momentum ${\mathbf{k}}_2$ and  $-{\mathbf{k}}_2$,
respectively. The momentum of pump laser for A is  ${\mathbf{k}}_1$, while that
for B is $-{\mathbf{k}}_1$. Thus, the mode $\mathbf{q}$ and $-\mathbf{q}$  are
in Bragg resonance with A  and B, respectively.  The output from the two phase
sensitive detectors (PSD) can be integrated by an integrator followed by
another  phase-sensitive detection of the integrated signal.}
\label{fig4}
\end{figure}
 $\hat{c}_{probe,B}^{(out)}$ is given by similar expression as above with
 $\alpha$ replaced  by $\beta$; and the subscripts  A and $\mathbf{q}$ replaced
 B and $-\mathbf{q}$, respectively. Thus  the output probes have oscillating
 parts at frequency $\delta - \omega_q^B$ proportional to the quasiparticle
 amplitudes $\alpha_{\mathbf{q}}$ and $\beta_{-\mathbf{q}}$, and at frequency
 $\delta + \omega_q^B$ proportional to the amplitues $\alpha_{-\mathbf{q}}$ and
 $\beta_{\mathbf{q}}$. Therefore, phase-senitive measurements of the spectral
 components of the output probe beams corresponding to these frequencies would
 provide  measures of the quasiparticle operators. The output from both the
 PSDs can be integrated  by an integrator and the integrated signal can also be
 measured by another PSD. By repeating the same measurements under identical
 conditions we could  calculate the number variances or correlation functions
 of interest, which can be employed to calculate the entanglement parameter in
 number operators, i.e., $\xi_n$. For calculating entanglement parameters in
 quadrature phase variables, both the output probe beams coming  from A and B,
 can be mixed via a beam splitter to form the superposition operators $\Sigma$
 which can be measured by a similar phase sensitive detection scheme (not shown
 in Fig.4). 
  
  Following the recent experiment of Ketterle's group \cite{ketter}, we also
  suggest that the quasiparticles can be detected by imparting a large momentum
  to them with additional Bragg pulses. Alternatively, in the large momentum
  regime ($q >\!> \xi^{-1}$), the Bragg-scattered atoms which essentially
  behave as free particles ($\omega_q^B \propto q^2$), can be outcoupled by
  swithcing off the trap. Since entanglement is between two opposite momentum
  states, by proper geometric arrangement, the two moving entangled atomic
  ensembles can be made to collide and interfere. From the interference pattern
  obtainable via absorption imaging, the atomic number fluctuations can be
  deduced using the theoretical model used in Ref.\cite{orzel}, and thus
  entanglement parameter in number variables can be calculated. 

   In conclusion, we have theoretically demonstrated how light scattering 
   leads to quantum entanglement between two Bose-Einstein condensates. We find
   that  the quasiparticle or phonon as well as free-particle momentum modes of
   two condensates can be  entangled.  The quasiparticle state can be
   sufficiently long-lived due to weak nature of interatomic interaction and
   the constraints imposed by  momentum conservations. The generated
   entanglement may be  useful in quantum communication using coherent light
   \cite{duan}. We have particularly focused on the conditions under which the
   entanglement can be obtained. We have also suggested how quasiparticles 
   could be studied by using Bragg scattering of far off resonant fields.

\end{document}